\documentclass[longauth]{aa}  

\usepackage{graphicx}
\usepackage{txfonts}
\usepackage{color}
\newcommand*\grg{J1235$+$5317}
\newcommand*\rad{~rad\,m$^{-2}$}
\newcommand*\cc{~cm$^{-3}$}
\newcommand*\sigmaRM{\sigma_{\rm RM}}

\begin{document} 

   \title{The intergalactic magnetic field probed by a giant radio galaxy}


   \author{S.~P.~O'Sullivan
          \inst{1}
          \and J.~Machalski
          \inst{2}
          \and C.~L.~Van Eck
          \inst{3}
          \and G.~Heald
          \inst{4}
          \and M.~Br\"uggen
          \inst{1}
          \and J.~P.~U.~Fynbo
          \inst{5}
          \and K.~E.~Heintz
          \inst{6}
          \and M.~A.~Lara-Lopez
          \inst{7}
          \and V.~Vacca
          \inst{8}
          \and M.~J.~Hardcastle
          \inst{9}
          \and T.~W.~Shimwell
          \inst{10, 11}
          \and C.~Tasse
          \inst{12, 13}        
          \and F.~Vazza
          \inst{14,1,19}  
          \and H.~Andernach
          \inst{15}
          \and M.~Birkinshaw
          \inst{16}
          \and M.~Haverkorn
          \inst{17}
          \and C.~Horellou
          \inst{18}
          \and W.~L.~Williams
          \inst{9}
          \and J.~J.~Harwood
          \inst{9}
          \and G.~Brunetti
          \inst{19}
          \and J.~M.~Anderson 
          \inst{20}
          \and S.~A.~Mao
          \inst{21}
          \and B.~Nikiel-Wroczy\'nski
          \inst{2}
          \and K.~Takahashi
          \inst{22}
          \and E.~Carretti
          \inst{19} 
          \and T.~Vernstrom
          \inst{23}
          \and R.~J.~van Weeren
          \inst{11}
          \and E.~Orr\'u
          \inst{10}
          \and L.~K.~Morabito
          \inst{24}
          \and J.~R.~Callingham
          \inst{10}
          }

   \institute{Hamburger Sternwarte, Universit\"at Hamburg, Gojenbergsweg 112, D-21029 Hamburg, Germany.\\
              \email{shane@hs.uni-hamburg.de}
         \and
             Astronomical Observatory, Jagiellonian University, ul. Orla 171, Krak\'ow PL 30-244, Poland.
         \and
         Department of Physics and Astronomy, University of Calgary, Calgary, Alberta, T2N 1N4, Canada.
         \and
         CSIRO Astronomy and Space Science, PO Box 1130, Bentley WA 6102, Australia.
         \and
         The Cosmic Dawn Center, Niels Bohr Institute, University of Copenhagen, Juliane Maries Vej 30, DK-2100 Copenhagen \O, Denmark 
         \and
         Centre for Astrophysics and Cosmology, Science Institute, University of Iceland, Dunhagi 5, 107 Reykjav\'ik, Iceland 
         \and
         Dark Cosmology Centre, Niels Bohr Institute, University of Copenhagen, Juliane Maries Vej 30, DK-2100 Copenhagen \O, Denmark 
         \and
         INAF - Osservatorio Astronomico di Cagliari, Via della Scienza 5, I-09047 Selargius (CA), Italy 
         \and
         Centre for Astrophysics Research, School of Physics, Astronomy and Mathematics, University of Hertfordshire, College Lane, Hatfield AL10 9AB, UK
         \and
         ASTRON, the Netherlands Institute for Radio Astronomy, Postbus 2, 7990 AA, Dwingeloo, The Netherlands
         \and
         Leiden Observatory, Leiden University, PO Box 9513, NL-2300 RA Leiden, The Netherlands
         \and 
         GEPI \& USN, Observatoire de Paris, Universit\'e PSL, CNRS, 5 Place Jules Janssen, 92190 Meudon, France 
         \and 
         Department of Physics \& Electronics, Rhodes University, PO Box 94, Grahamstown, 6140, South Africa
         \and
         Dipartimento di Fisica e Astronomia, Universit\'a di Bologna, Via Gobetti 93/2, 40121, Bologna, Italy
         \and
         Departamento de Astronom\'ia, DCNE, Universidad de Guanajuato, Guanajuato, Mexico
         \and 
         HH Wills Physics Laboratory, University of Bristol, Tyndall Avenue, Bristol, BS8 1TL, UK
         \and 
         Department of Astrophysics/IMAPP, Radboud University Nijmegen, PO Box 9010, 6500 GL Nijmegen, the Netherlands
         \and 
         Dept. of Space, Earth and Environment, Chalmers University of Technology, Onsala Space Observatory, SE-43992 Onsala, Sweden
         \and
         INAF - Istituto di Radioastronomia, via P. Gobetti 101, 40129, Bologna, Italy
         \and
         GFZ German Research Centre for Geosciences, Telegrafenberg, 14473 Potsdam, Germany     
         \and 
         Max-Planck-Institut f\"ur Radioastronomie, Auf dem H\"ugel 69, D-53121 Bonn, Germany
         \and
         Department of Physics, Kumamoto University, Kumamoto 860-8555, Japan
         \and
         Dunlap Institute for Astronomy and Astrophysics University of Toronto, Toronto, ON M5S 3H4, Canada
         \and
         Astrophysics, University of Oxford, Denys Wilkinson Building, Keble Road, Oxford OX1 3RH, UK
             }

   \date{Received 12 July, 2018; accepted 9 October, 2018}

 
\abstract{Cosmological simulations predict that an intergalactic magnetic field (IGMF) pervades the large scale structure (LSS) of the Universe. Measuring the IGMF is important to determine its origin (i.e.~primordial or otherwise). 
Using data from the LOFAR Two Metre Sky Survey (LoTSS), we present the Faraday rotation measure (RM) and depolarisation properties of the giant radio galaxy \grg, at a redshift of $z=0.34$ and 3.38~Mpc in size. 
We find a mean RM difference between the lobes of $2.5\pm0.1$\rad, in addition to small scale RM variations of $\sim 0.1$\rad. From a catalogue of LSS filaments based on optical spectroscopic observations in the local universe, we find an excess of filaments intersecting the line of sight to only one of the lobes. 
Associating the entire RM difference to these LSS filaments leads to a gas density-weighted IGMF strength of $\sim$0.3~$\mu$G. However, direct comparison with cosmological simulations of the RM contribution from LSS filaments gives a low probability ($\sim$5\%) for an RM contribution as large as 2.5\rad, for the case of IGMF strengths of 10 to 50 nG. 
It is likely that variations in the RM from the Milky Way (on 11\arcmin~scales) contribute significantly to the mean RM difference, and a denser RM grid is required to better constrain this contribution. 
In general, this work demonstrates the potential of the LOFAR telescope to probe the weak signature of the IGMF. 
Future studies, with thousands of sources with high accuracy RMs from LoTSS, will enable more stringent constraints on the nature of the IGMF. 
}


   \maketitle
%

\section{Introduction}\label{sec:introduction}
Diffuse gas is expected to permeate the large-scale structure (LSS) of the Universe away from galaxy groups and clusters. Detecting and characterising this intergalactic gas is challenging due to the expected low particle number density ($\sim$$10^{-5}$ to $10^{-6}$~\cc) and temperature ($10^5$ to $10^7$~K). Although diffuse, this warm-hot intergalactic medium \citep[WHIM;][]{dave2001,cenostriker2006} potentially contains half the total baryon content of the local Universe \citep{bregman2007,nicastro2018}. In addition, accretion shocks along these LSS filaments are predicted to accelerate particles to relativistic energies and to amplify magnetic fields. Thus, detecting this filamentary structure in synchrotron emission using radio telescopes is a promising avenue for studying the WHIM \citep[e.g.][]{vazza2015a}. 
Recent statistical studies based on the cross-correlation of diffuse radio synchrotron emission and the underlying galaxy distribution have derived upper limits on the magnetisation of filaments of the order of $0.1$~$\mu$G \citep{vernstrom2017,brown2017}. 
Furthermore, \cite{vacca2018} found a faint population of sources which might be the tip of the iceberg of a class of diffuse large-scale synchrotron sources associated with the WHIM connected to a large-scale filament of the cosmic web.
An alternative approach is to measure the Faraday rotation properties of the magnetised WHIM using many bright, polarised, background radio sources \citep[e.g.][]{stasyszyn2010,akahori2014,vacca2016}. 

From simulations, the field strength of the intergalactic magnetic field (IGMF) is expected to be in the range of 1 to 100 nG \citep[e.g.][]{dolag1999, brueggen2005, ryu2008,vazza2017}. It is important to constrain the magnetic field in the WHIM in order to determine the unknown origin of the large scale magnetic field in the Universe \citep{zweibel2006}. 
While large scale fields are commonly detected in galaxies and galaxy clusters, the strong modification of these fields erases the signature of their origin \citep[e.g.][]{vazza2015b}. 
This may not be the case in the WHIM, as the amplification of primordial magnetic fields in these filamentary regions are likely primarily due to compressive and shearing gas motions, in addition to small-scale shocks, such that the observed level of magnetisation could be connected to the seeding process \citep[e.g.][]{ryu2008,vazza2014}. 
The AGN and star formation activity in galaxies can also drive powerful outflows that may significantly magnetise the intergalactic medium on large scales \citep[e.g.][]{furlanettoloeb2001,donnert2009,beck2013}. Therefore, distinguishing between a primordial origin and a later injection of magnetic field that was initially generated on smaller scales by galaxies and stars is a key goal for studies of the IGMF \citep[see][and references therein]{akahori2018}. 

It has also been proposed to study the WHIM using large or `giant' radio galaxies (GRGs) whose linear size can extend beyond 1 Mpc, with the largest such example being 4.7 Mpc in extent \citep{machalski2008}. GRGs are usually FRII type radio galaxies \citep[e.g.][]{dabhade2017}, although some giant FRI also exist \citep[e.g.][]{heesen2018,horellou2018}, that extend well beyond their host galaxy and local environments, into the surrounding intergalactic medium. Asymmetries in the GRG morphology can be used as a probe of the ambient gas density \citep{subrahmanyan2008,safouris2009,pirya2012,malarecki2015} and the Faraday rotation properties of the polarised emission from the lobes can be used to study the magnetic field properties of the surrounding gas on Mpc scales \citep{xu2006,osullivanlenc2018}. 
Another potential approach to studying the magnetised WHIM in cluster outskirts is by using Faraday rotation observations of the highly polarised emission from radio relics \citep[e.g.][]{kierdorf2017, loi2017}. 

The effect of Faraday rotation is measured through its influence on the linear polarisation vector as a function of wavelength-squared. 
The observed Faraday rotation measure, RM [rad~m$^{-2}$], depends on the line-of-sight magnetic field, $B_{||}$~[$\mu$G], threading a region of ionised gas with electron density, $n_{\rm e}$~[cm$^{-3}$], along a path length, $l$ [pc], following
\begin{equation}
\label{eqn:rm}
\mathrm{RM} = 0.812\int_{\rm source}^{\rm telescope} n_{\rm e} \, B_\parallel \, \mathrm{d}l  \,\,\,\,\, {\rm rad~m}^{-2}.
\end{equation}   

\noindent In this paper, we present an analysis of the linear polarisation and Faraday rotation properties of an FRII radio galaxy (\grg) with a linear size of 3.4~Mpc. The observations were done with the Low Frequency Array \citep[LOFAR;][]{vanhaarlem2013} which provides excellent sensitivity to diffuse extended structures due to the presence of numerous short baselines and exceptional Faraday rotation measure (RM) accuracy, which depends on the total coverage in wavelength-squared. While low frequency radio telescopes provide the best RM accuracy, sources at these frequencies are most strongly affected by Faraday depolarisation \citep[e.g.][]{burn1966}, which decreases the degree of linear polarisation below the detection limit for many sources \citep[][]{farnsworth2011}. Despite this there is a growing number of polarised sources being found at low frequencies \citep[e.g.][]{Bernardi:2013, Mulcahy:2014, Jelic:2015, orru2015, Lenc:2016, vaneck2018, osullivanlenc2018,neld2018,riseley2018}.

\grg~was discovered to be polarised at 144~MHz by \cite{vaneck2018}, in LOFAR data imaged at an angular resolution of 4.3\arcmin. 
The source was first reported by \cite{schoenmakers2001}, and the first optical identification (SDSS J123458.46$+$531851.3) was proposed by \cite{banfield2015}. 
However, our new observations show that the previously
assumed host galaxy is accidentally located close to the geometric
centre between the two lobes and that the real host galaxy is
actually connected to the south east (SE) lobe by a faint jet.
The radio core is coincident with the galaxy SDSS~J123501.52$+$531755.0, which is 
identified as PSO~J123501.519$+$531754.911 \citep{flewelling2016} for the radio source ILT~J123459.82$+$531851.0 in \cite{williams2018}.
Estimates of the photometric redshift of this galaxy are 0.349 \citep{bilicki2016}, 0.41 \citep{beckdobos2016} and 0.44 \citep{brescia2014,duncan2018}.

The host galaxy is identified in \cite{hao2010} as a red-sequence galaxy and a cluster candidate, GMBCG~J188.75636+53.29864. 
This is intriguing as GRGs are often thought to evolve in underdense galaxy environments \citep[e.g.][]{mack1998}, 
however, recent work indicates that they are most likely the oldest sources in the general population of powerful radio galaxies \citep{hardcastle2018}. 
In addition, \cite{hao2010} estimate a total of $\sim$9 galaxies within 0.5~Mpc with luminosities $L > 0.4 L^*$, using a weak-lensing 
scaling relation, which suggests a poor cluster environment. 
There is also no evidence for a massive cluster at this location in the sky in the Planck thermal 
Sunyaev-Zeldovich map \citep{planckymap}. 

This paper presents a follow-up study using the same LOFAR data as \cite{vaneck2018}, but 
imaging at higher angular resolution. 
We also confirm the new optical host identification and determine its spectroscopic redshift as $z\sim0.34$, giving the 
projected linear size of 3.4~Mpc.
In Section~\ref{sec:obs}, we describe the radio polarisation and optical spectroscopic observations. 
Section~\ref{sec:results} presents the physical properties of \grg, the inference on the properties of 
its environment based on dynamical modelling of the jets, and the RM and depolarisation behaviour. 
In Section~\ref{sec:discuss} we discuss the results in the context of the study of the intergalactic 
medium and its magnetisation. The conclusions are listed in Section~\ref{sec:conclusion}. 
Throughout this paper, we assume a $\Lambda$CDM cosmology with 
H$_0 = 67.8$ km s$^{-1}$ Mpc$^{-1}$, $\Omega_M=0.308$ and $\Omega_{\Lambda}=0.692$ \citep{planck2016xiii}.
At the redshift of the source, 1\arcsec~corresponds to a linear size of 5.04~kpc. 
We define the total intensity spectral index, $\alpha$, 
such that the observed total intensity ($I$) at frequency $\nu$ follows the relation 
$I_{\nu}\propto\nu^{\rm{+}\alpha}$.

\section{Observations \& Data Analysis}
\label{sec:obs}

\subsection{Radio observations}
The target source \grg~was observed as part of the LOFAR Two-Metre Sky Survey \citep[LoTSS;][]{shimwell2017,shimwell2018}, 
which is observing the whole northern sky with the LOFAR High-Band Antenna (HBA) from 120 to 168 MHz.  
The data relevant to our target were observed in full polarisation for 8 hours on 26 June 2014, as part of the observing program LC2\_038 
and with a pointing centre of J2000 12$^{\rm{h}}$38$^{\rm{m}}$06$\fs$7, $+52$\degr07$\arcmin$19$\arcsec$. 
This gives a distance of $\sim$1.26\degr~of the target \grg~from the pointing centre (the FWHM of the primary beam is $\sim$4\degr). 
Direction-independent calibration was performed using the {\sc prefactor} pipeline\footnote{https://github.com/lofar-astron/prefactor}, 
as described in detail in \cite{shimwell2017} and \cite{degasperin2018}, which includes the ionospheric RM correction using 
{\sc rmextract}\footnote{https://github.com/lofar-astron/RMextract}. 
Residual ionospheric RM correction errors of $\sim$0.05\rad~are estimated between observations \citep{vaneck2018}, while slightly larger errors of $\sim$0.1 to 0.3\rad~are estimated across a single 8-hour observation \citep{sotomayor2013}. 

The resulting measurement set, after the {\sc prefactor} pipeline, has a time resolution of 8~s and a frequency resolution of 97.6~kHz. 
The direction-independent calibrated data are used throughout for the polarisation and rotation measure analysis, while the direction-dependent calibrated total intensity image \citep{shimwell2018} is used to determine the source morphological properties with high precision and for the identification of the host galaxy location. 
Analysis of polarisation and rotation measure data products after direction-dependent calibration will be presented in future work. 

\subsection{Polarisation and Faraday rotation imaging}
\label{sec:rmspecs}
To analyse the polarisation and Faraday rotation properties of the target, we phase-shifted the calibrated uv-data 
to the coordinates of the host galaxy (12$^{\rm{h}}$35$^{\rm{m}}$01$\fs$5, $+53$\degr 17$\arcmin$55$\arcsec$), which 
lies almost at the centre of the extended emission. 
We calibrated the data for short-timescale phase variations caused by the ionosphere, then averaged to 32 s to reduce the data size 
and to help speed up the subsequent imaging, while avoiding any significant time smearing \citep[e.g.][]{neld2018}. 
Both the phase-shifting and time-averaging were done using NDPPP \citep{2018ascl.soft04003V}\footnote{https://support.astron.nl/LOFARImagingCookbook/}. 
The imaging software {\sc wsclean} \citep{offringa-wsclean-2014}\footnote{https://sourceforge.net/projects/wsclean} 
was used to create $I$, $Q$, $U$, $V$ channel images at 
97.6~kHz resolution, for a 25\arcmin~field of view ($\sim$twice the linear size of \grg). 
A minimum uv-range of 150~$\lambda$ was used to avoid sensitivity to Galactic polarised emission on scales of $\gtrsim25\arcmin$. 
The maximum uv-range was set to 18~k$\lambda$, and combined with a Briggs weighting of 0, 
resulted in a beam size of $26\arcsec \times 18\arcsec$, sampled with $3\arcsec \times 3\arcsec$~pixels. 
The differential beam correction per channel was applied using {\sc wsclean}, as the correction for the LOFAR beam gain at 
the pointing centre was already applied during the initial calibration of the data. 
All channel images with $Q$ or $U$ noise higher than five times the average noise level were removed 
from subsequent analysis, leaving a total of 404 images covering 120 to 167 MHz (with a central frequency of 143.5 MHz).  

RM synthesis and {\sc rmclean} \citep{bdb2005,heald2009} were then applied to the $Q$ and $U$ images using {\sc pyrmsynth}\footnote{https://github.com/mrbell/pyrmsynth}. 
The data have an RM resolution of 1.16\rad, are sensitive to polarised emission from Faraday thick regions up to $\sim$0.98\rad, and $|$RM$|$ values for Faraday thin regions as high as 450\rad~can be detected. An RM cube with a Faraday depth ($\phi$) axis covering $\pm500$\rad~and sampled at 0.5\rad~intervals was constructed for initial inspection of the data. 
The concept of Faraday depth \citep{burn1966} can be useful to introduce here to describe regions with complicated distributions of Faraday rotation along the line of sight, such as multiple distinct regions of polarised emission experiencing different amounts of Faraday rotation, which could be identified through multiple peaks in a Faraday depth spectrum or Faraday dispersion function (FDF). 
As no significant emission was found at large Faraday depths, the final RM and polarisation images were 
constructed from FDFs with a range of $\pm150$\rad, sampled at 0.15\rad. 
To identify peaks in the FDF, a threshold of 8$\sigma_{QU}$ was used, where $\sigma_{QU}$ is calculated from the outer 20\% of the Faraday depth range in the {\sc rmclean} $Q$ and $U$ spectra. The mean $\sigma_{QU}$ across the field was $\sim$90~$\mu$Jy~beam$^{-1}$. 
Since no correction was made for the instrumental polarisation, peaks in the Faraday dispersion function appears near $\phi\sim0$\rad~at a typical level of $\sim$1.5\% of the Stokes $I$ emission. This instrumental polarisation signal is also smeared out by the ionospheric RM correction making it difficult to identify real polarised emission at low Faraday depths ($\lesssim\pm3$\rad). Thus, when identifying real polarised emission peaks in the FDF, the range $\pm3$\rad~is excluded. RM and polarised intensity images are created from the brightest, real polarised peak above 8$\sigma_{QU}$ at each pixel, after fitting a parabola around the peak to obtain the best-fitting RM and polarised intensity. In the case of the polarised intensity image, a correction for the polarisation bias was also made following \cite{george2012}. The error in the RM at each pixel was calculated in the standard way as the RM resolution divided by twice the signal to noise ratio of the detection \citep{bdb2005}. 

A full-band Stokes $I$ image was made using the same image parameters as the channel images specified above, with multi-scale cleaning applied for an automatic threshold of 3$\sigma$ and deeper cleaning (to 0.3$\sigma$) within an automatic masked region created from the clean components. The degree-of-polarisation image was created by dividing the band-averaged polarised intensity image from RM synthesis (with a cutoff at 8$\sigma_{QU}$) by the full-band Stokes $I$ image (with a cutoff at 3 times the local noise level). 

\begin{figure}
\begin{center}
\includegraphics[width=0.98\linewidth]{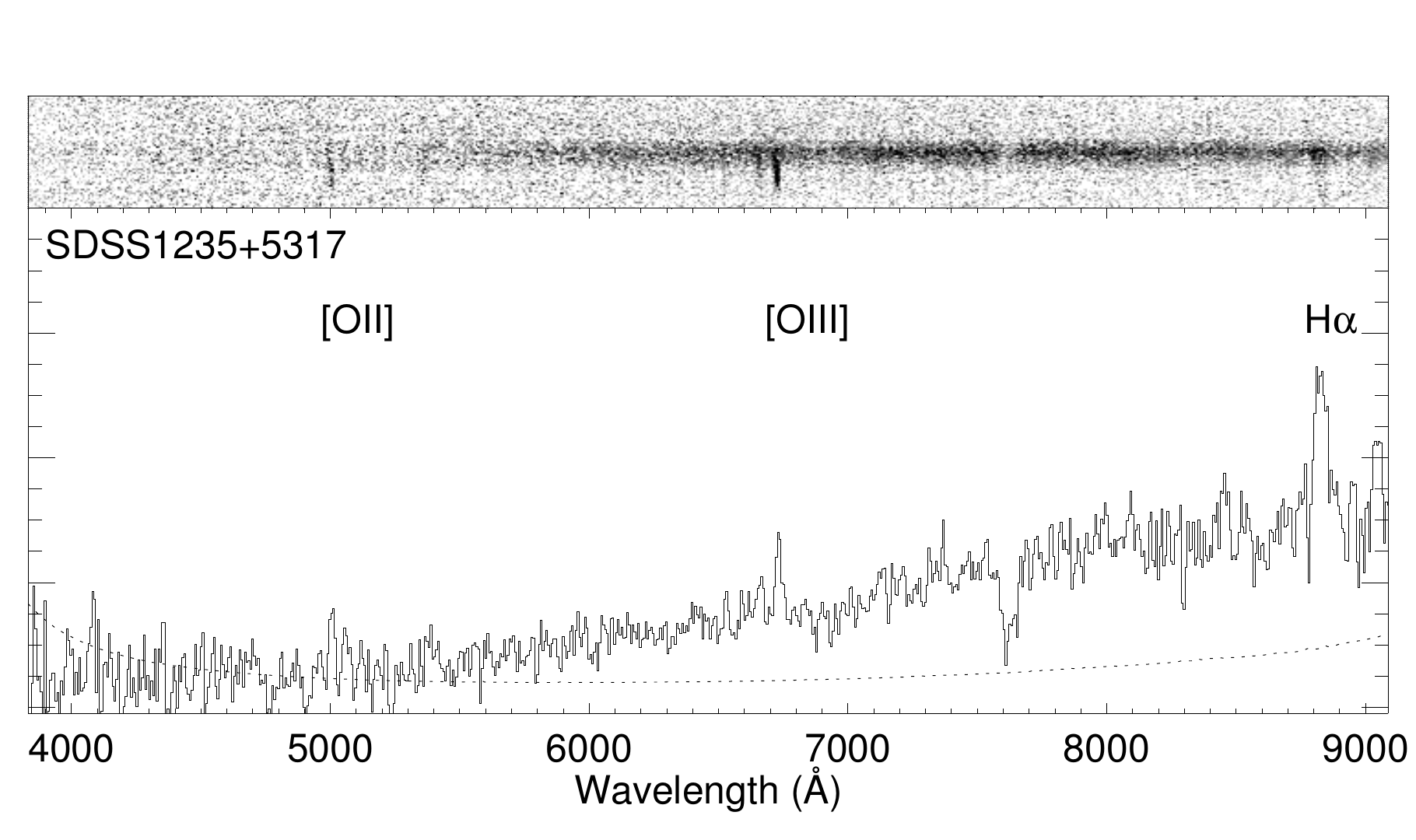}
\caption{Optical spectrum of the host galaxy SDSS~J123501.52$+$531755.0 taken with
AIFOSC instrument on the Nordic Optical Telescope, which shows 
emission lines H$\alpha$, [O{\sc ii}] and [O{\sc iii}] at a redshift of 0.34. }
\label{fig:redshift}
\end{center}
\end{figure}

\begin{figure}
\begin{center}
\includegraphics[width=1\columnwidth,clip=true,trim=0.0cm 0.0cm 1.0cm 0.0cm]{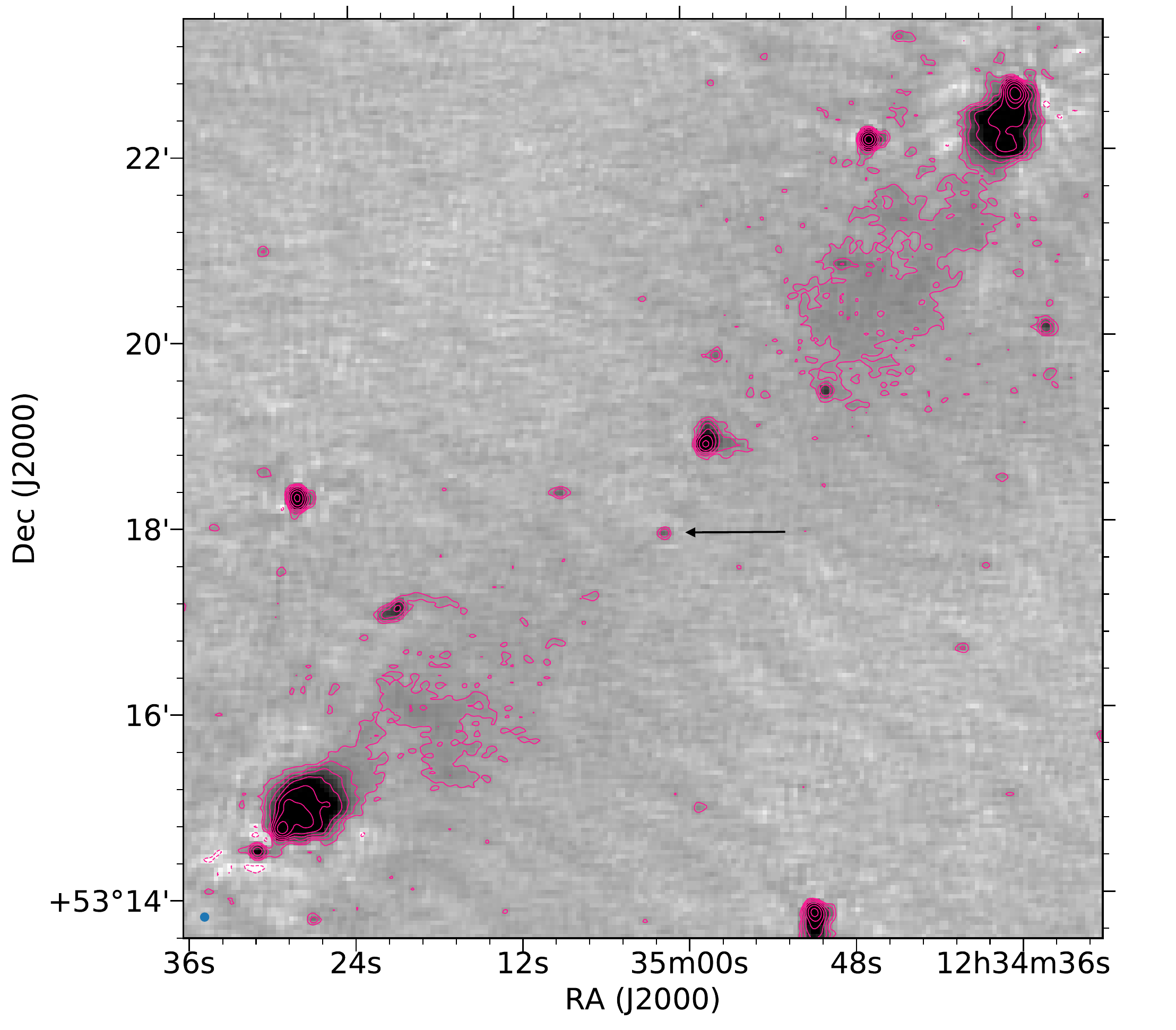}
\caption{LoTSS total intensity image at 144 MHz at 6\arcsec~resolution (after direction-dependent calibration). 
The contours start at 300~$\mu$Jy~beam$^{-1}$ and increase by factors of 2 (with one negative contour at $-300$~$\mu$Jy~beam$^{-1}$). 
The greyscale image is tuned to show the noise variation across the image ($\sim$70~$\mu$Jy~beam$^{-1}$ away from bright sources 
and $\sim$100~$\mu$Jy~beam$^{-1}$ near the hotspots), as well as a faint hint of the south-east jet. The radio galaxy 
core coincident with the host galaxy SDSS~J123501.52$+$531755.0 is indicated by the horizontal arrow. 
The synthesised beam size is shown in the bottom left hand corner of image.}
\label{fig:IDDF}
\end{center}
\end{figure}

\subsection{Optical spectroscopic observations}
SDSS~J123501.52$+$531755.0 was observed with the Nordic Optical Telescope on March 25 and 
March 26 2018 for a total integration time of 5400 sec. We used the Andalucia Faint 
Object Spectrograph and Camera (AlFOSC) and a 1.3 arcsec wide longslit and 
grism 4 with 300 rules per millimetre providing a spectral resolution of 280 and a useful 
spectral range of 3800 to 9100~\AA. The slit was placed at a parallactic angle  
of 60 degrees east of north on both nights at the onset of integration. 
The airmass ranged from 1.20 to 1.15. The observing conditions were poor with a variable 
seeing above 2 arcsec and with passing clouds. Despite this we clearly detected several 
emission lines (Figure~\ref{fig:redshift}) consistent with a mean redshift of $0.3448\pm0.0003$ (1-sigma error). 
The [O{\sc ii}] and [O{\sc iii}] images have a peculiar morphology 
extending away from the continuum source to the northern side of the galaxy. In particular 
[O{\sc iii}],$\lambda$5008\,\AA~can be traced over 4 arcseconds below the continuum trace (20 kpc at $z=0.34$). 
This indicates the presence of an extended emission line region. 

\begin{figure*}
\begin{center}
\includegraphics[width=0.49\linewidth]{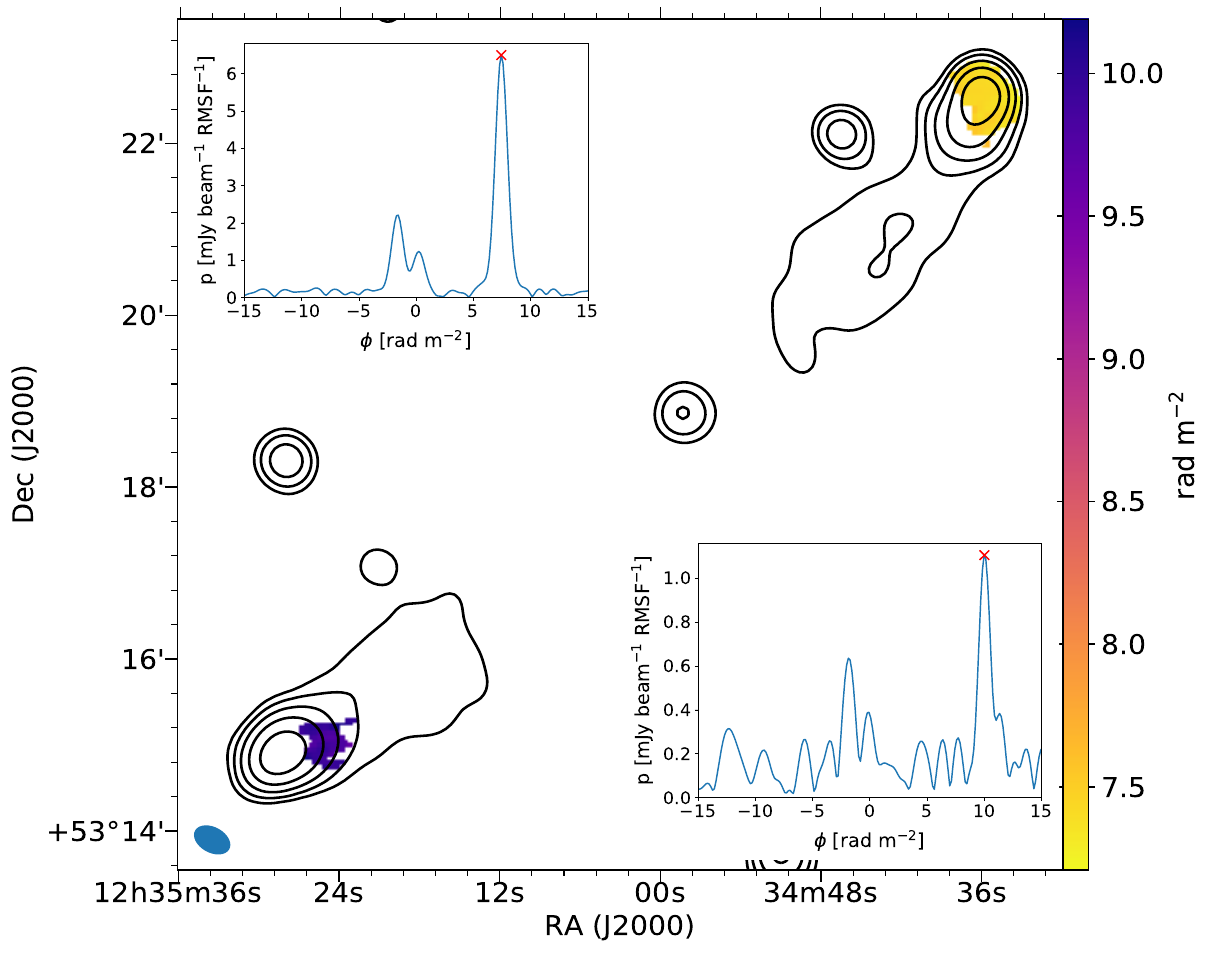}
\includegraphics[width=0.49\linewidth]{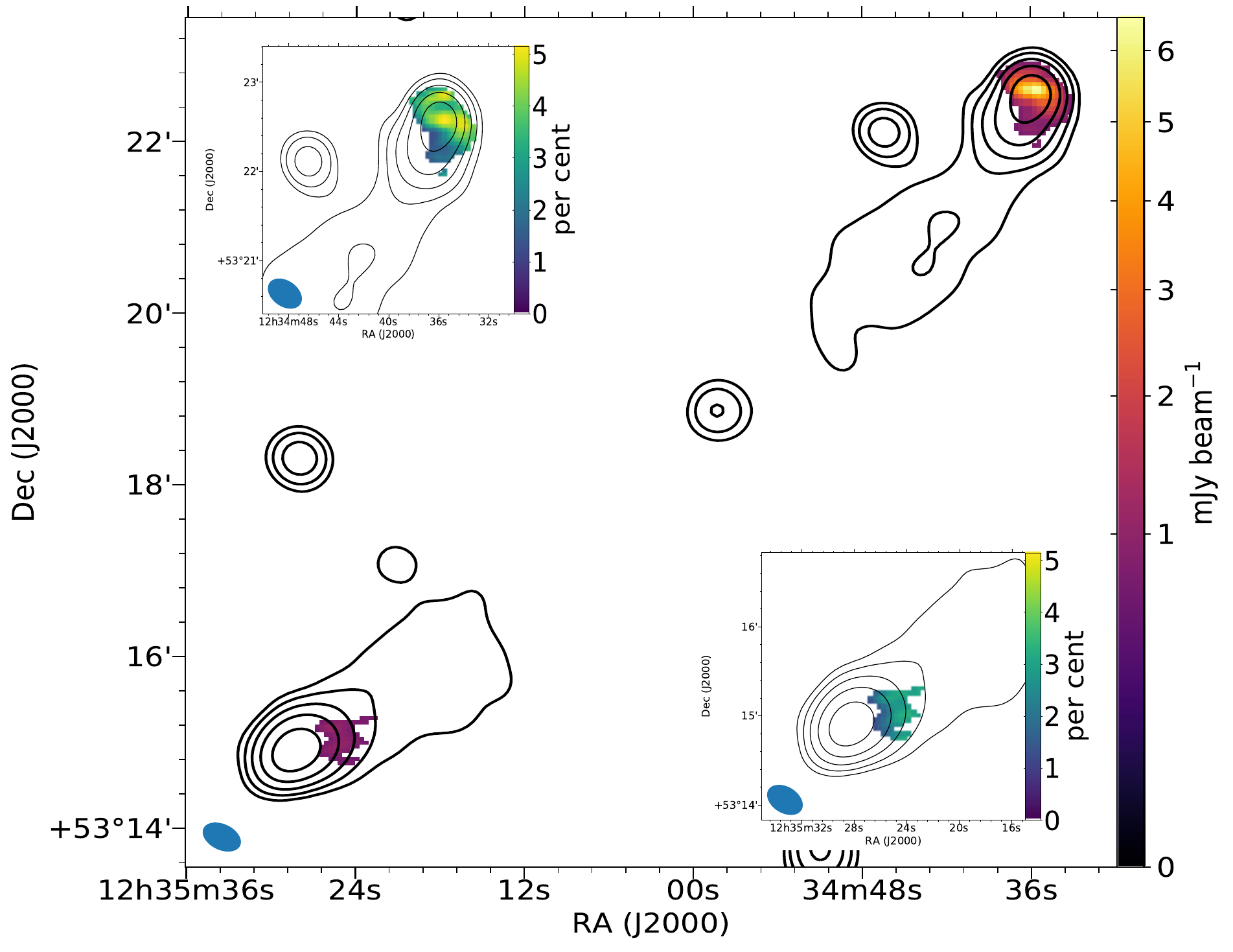}
\caption{Left image: Main image: Faraday rotation measure distribution (colour scale) of the 
north-west (NW) and south-east (SE) lobe regions that are detected above the threshold of $8\sigma_{QU}$, 
overlaid by the total intensity contours starting at 5~mJy~beam$^{-1}$ and increasing in factors of two.
Insets: The absolute value of the {\sc rmclean} Faraday dispersion function for the brightest polarised 
pixel in the NW lobe (top) and SE lobe (bottom).
Right image: Main image: polarized intensity greyscale, in mJy~beam$^{-1}$, overlaid by the total intensity contours. 
Insets: degree of polarisation colourscale (in per cent) from zoomed in regions of the NW and SE lobes.}
\label{fig:RMfpol}
\end{center}
\end{figure*}

\section{Results}
\label{sec:results}

\subsection{Radio morphology of \grg}\label{sec:stokesi}
Figure~\ref{fig:IDDF} shows the total intensity image at 6\arcsec~resolution from the LoTSS direction-dependent calibrated data \citep{shimwell2018}. This provides the best radio image to date for this source, enabling an unambiguous host galaxy identification with SDSS~J123501.52$+$531755.0.
The noise level in the image ranges from $\sim$70~$\mu$Jy~beam$^{-1}$ in areas away from bright sources to $\sim$100~$\mu$Jy~beam$^{-1}$ near the hotspots/lobes. 

The core of this FRII radio galaxy, located at J2000 12$^{\rm{h}}$35$^{\rm{m}}$01$\fs$5, $+53$\degr 17$\arcmin$55$\arcsec$, has an integrated flux density of $\sim$1.1~mJy at 144 MHz and 1.4 GHz \citep[FIRST;][]{becker1995} suggesting a flat spectrum. However, the core is also detected in the VLASS\footnote{https://archive-new.nrao.edu/vlass/} Quick-Look (QL) image at 3~GHz ($\sim$2.9~mJy) and the 9C catalogue \citep{waldram2010} at 15 GHz ($\sim$4~mJy) indicating an inverted spectral index of $\alpha_{\rm core}\sim+0.3$ when combined with the LoTSS core flux density. As the LoTSS, VLASS and 9C observations are closest in time, we consider the core to have an inverted spectral index, with time variability explaining the lower than expected flux density from FIRST at 1.4~GHz. 
There is also a faint hint of a jet connecting the host with the south-east (SE) lobe. If this is real, then it suggests that the SE jet and lobe are orientated slightly towards us on the sky.  

Using the $3\sigma$ contour to define the lobe edges, 
we find the lobes have a width of $\sim$83\arcsec~and $\sim$94\arcsec, giving an axial
ratio of $\sim$4.4 for the north-west (NW) lobe and $\sim$3.3 for the SE 
lobe, respectively. This is consistent with the typical axial ratios from
2 to 7 for the lobes of most (smaller) GRGs \citep[e.g.][]{machalski2006}.
In Table~\ref{tab:fluxes}, we compile the integrated flux densities of the NW and SE 
lobes and hotspots from both current and archival data. 
The integrated flux densities of the NW lobe and hotspot are slightly higher than the SE lobe and hotspot at 144~MHz, with both having spectral index values of $\alpha_{\rm lobe}\sim-0.8$. 
The NW hotspot is resolved into primary and secondary hotspot regions in the VLASS at 3 GHz ($2.4\arcsec\times2.1\arcsec$ beam), while the SE hotspot maintains a single component. 

The straight-line distance from the core to the NW hotspot is $\sim$365\arcsec~(1.84~Mpc), compared to $\sim$311\arcsec~(1.56~Mpc) from the core to the SE hotspot, giving a lobe length ratio of 1.17. 
The inferred jet-misalignment (from co-linearity) of $\sim13.6\degr$ is most likely due to bending of the NW and/or SE jets on large scales, as is sometimes observed in other FRII radio sources \citep{black1992}. 
We expect that the lobe-length asymmetry and jet-misalignment are caused by interactions between the jet and the external environment on large scales, as opposed to light travel time effects \citep{longair1979}. Asymmetries in the jet and lobe lengths of GRGs are often attributed to interactions with the large scale structure environment \citep{pirya2012,malarecki2015}. The advancing NW jet may be influenced by a nearby filament (see Section~\ref{sec:filaments} and the filament in the $z\sim0.335$ slice), although deeper optical spectroscopic observations would be required to determine whether or not this filament is indeed close enough in redshift to that of the host galaxy to have an influence. 

\subsection{Faraday rotation measure distribution}\label{sec:RM}
\begin{figure}
\begin{center}
\includegraphics[width=0.93\linewidth,clip=true,trim=0.5cm 0.0cm 1.0cm 1.0cm]{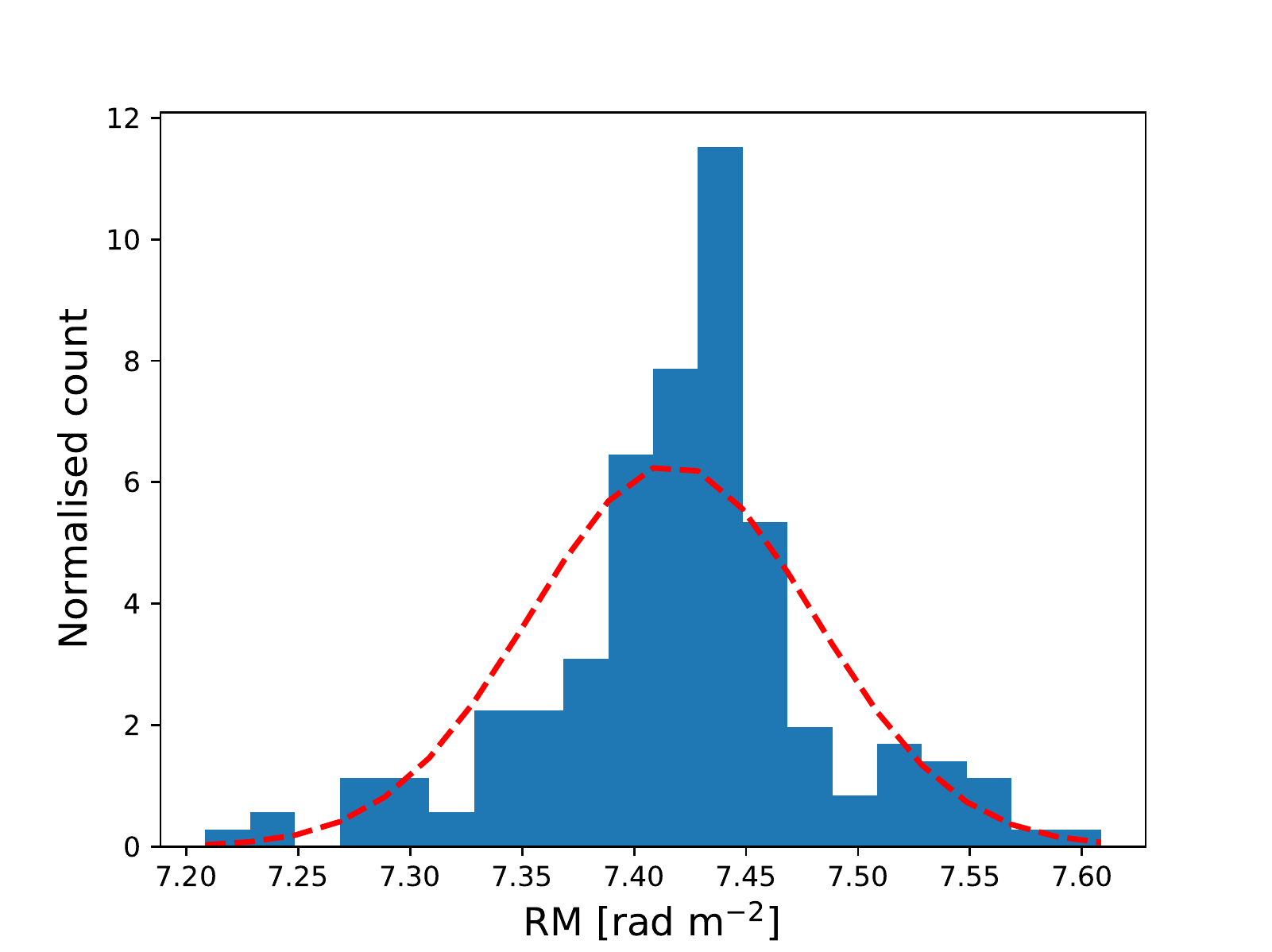}
\includegraphics[width=0.93\linewidth,clip=true,trim=0.5cm 0.0cm 1.0cm 1.0cm]{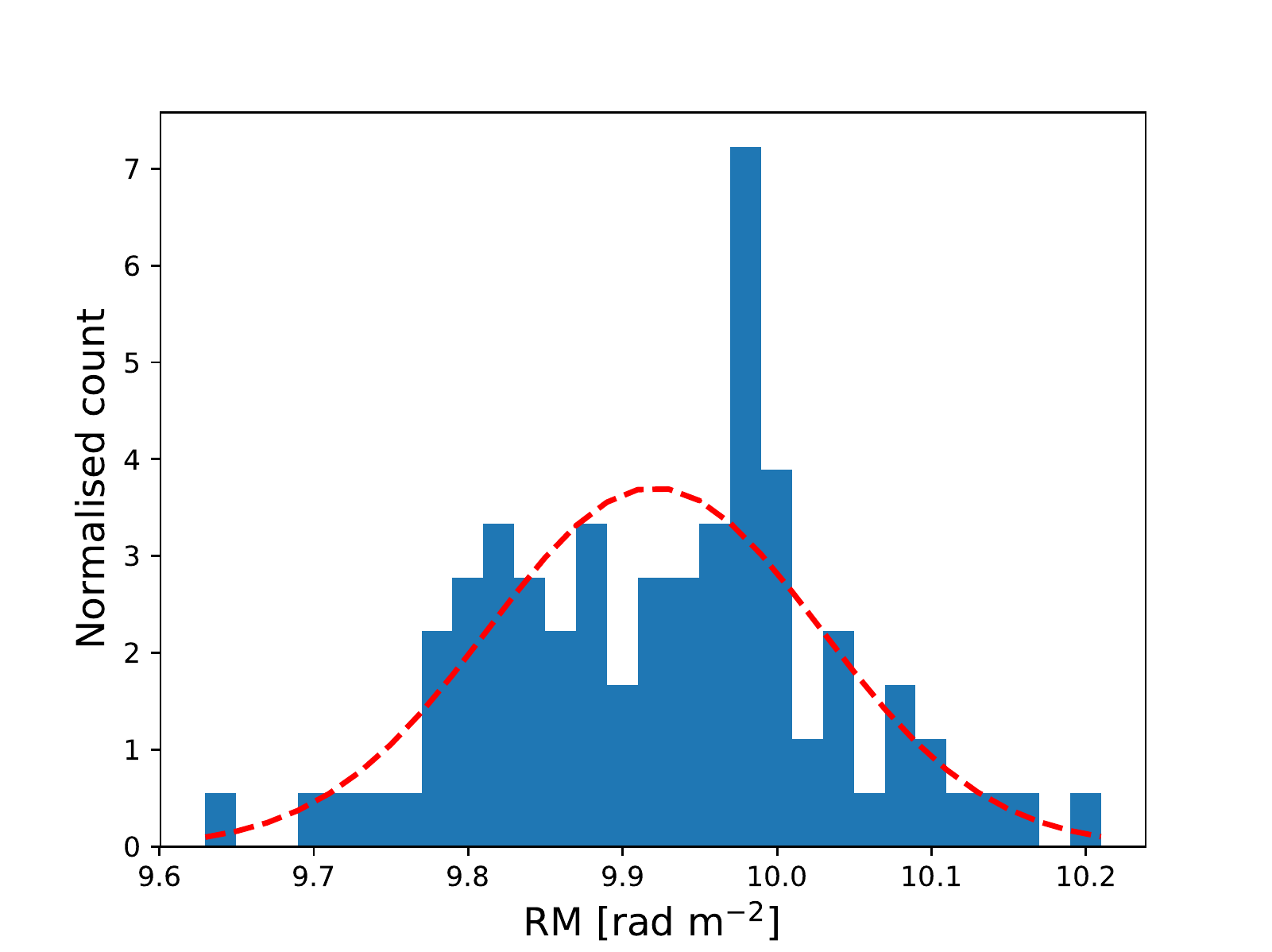}
\caption{Histograms of the RM distribution from the north-west lobe (top) and south-east lobe (bottom)  
regions of \grg. The red dashed line shows a Gaussian distribution with the same mean and 
standard deviation as the observed data. }
\label{fig:RMhist}
\end{center}
\end{figure}

Figure~\ref{fig:RMfpol} shows the RM distribution for \grg, using an $8\sigma_{QU}$ threshold, 
overlaid by Stokes $I$ contours at the same angular resolution. 
The Faraday dispersion functions for the brightest pixel in polarised intensity in each lobe are also shown, with 
a red cross marking the peak polarisation at which the RM was found. 
Other peaks in the spectrum are either noise peaks or related to the instrumental polarisation near RM~$\sim0$\rad. 
The RM distributions of each lobe are shown in Figure~\ref{fig:RMhist}. 
The mean and standard deviation of the RM are $+7.42$\rad~and 0.07\rad~for the NW lobe, 
and $+9.92$\rad~and 0.11\rad~for the SE lobe, respectively. The median RM errors for the NW and SE 
lobe regions are 0.04\rad~and 0.06\rad. 
The mean RM difference between the lobes of $2.5\pm0.1$\rad~is thus highly significant. 
At the angular separation of the lobes (11\arcmin), systematic errors in the ionospheric RM correction 
would affect both lobes equally and thus do not contribute to the RM difference between the lobes. 
We can estimate the significance of the small RM variations within each lobe 
accounting for the number of pixels in each synthesised beam following \cite{leahy1986}, 
where a reduced-chi-squared of $\sim$1 is expected 
if noise errors dominate the RM fluctuations. We find no evidence for the detection of significant RM variations across 
the NW lobe, with a reduced-chi-squared of 1.1. 
However, a reduced-chi-squared of 1.8 provides evidence, at a level of $\sim$1.35$\sigma$, for RM variations 
across the SE lobe of $\sim$0.1\rad. 

\subsection{Faraday depolarisation}\label{sec:depol}
The polarised intensity and degree of polarisation distributions are shown in Figure~\ref{fig:RMfpol}. 
The NW lobe is much brighter with a peak polarised intensity of 6.5~mJy~beam$^{-1}$ (coincident 
with the hotspot) and a 
degree of polarisation of 4.9\% at that location (ranging from 1.2\% to 5.1\% across the detected emission). 
The SE lobe is fainter with a peak polarised intensity of 1.1~mJy~beam$^{-1}$. The degree of polarisation 
at that location is 2.8\%, and it ranges from 1.1 to 3.3\% across the lobe. 
The non-detection of polarised emission from the SE hotspot is likely due to intrinsic non-uniform field 
structures and Faraday depolarisation on scales smaller than the resolution of our observations. 
The fainter, extended lobe emission would have to be $\gtrsim10\%$ polarised 
to be detected in these observations. 

In order to estimate the amount of depolarisation between 1.4 GHz and 144 MHz, the LoTSS data 
were compared with those of the NRAO VLA Sky Survey \citep[NVSS;][]{condon1998}. 
To determine the degree of polarisation at the same angular resolution as the NVSS survey, the 
RM pipeline was re-applied to the LoTSS data imaged at a lower angular resolution of $\sim$45\arcsec. 

At the peak polarised intensity location in the NW lobe of the LOFAR image, matched to the NVSS resolution, 
the degree of polarisation is $4.0\pm0.3$\%. At the same location in the NVSS image at 1.4 GHz, the degree 
of polarisation is $6.4\pm1.4$\%. This gives a depolarisation factor of ${\rm DP}_{1400}^{144}\sim0.6$, where 
${\rm DP}_{1400}^{144}$ is the degree of polarisation at 144 MHz divided by the degree of polarisation at 1.4 GHz. 
Assuming the commonly used external Faraday dispersion model for depolarisation, $p(\lambda)\propto{\rm e}^{-2\sigma_{\rm RM}^2\lambda^4}$ \citep{burn1966}, provides a value of $\sigmaRM\sim0.1$\rad. 

For the SE lobe, the degree of polarisation at the peak polarised intensity at 144~MHz is 
$1.8\pm0.7$\% (at 45\arcsec~resolution) and $10.1\pm2.1$\% at the same location at 1.4~GHz. 
This gives ${\rm DP}_{1400}^{144}\sim0.2$, corresponding 
to larger amounts of depolarisation than in the NW lobe. In the case of external Faraday dispersion, this corresponds 
to $\sigmaRM\sim0.2$\rad. 

The observed difference in depolarisation between the NW and SE lobes may be due to 
the different location within each lobe from which the polarised emission arises. In the case of the NW lobe, 
the peak polarised emission is coincident with the hotspot location, whereas in the SE lobe, 
the peak polarised emission is significantly offset from the hotspot ($\sim$40\arcsec~away, 
in the bridge emission, with the offset also present in the NVSS images). 
Furthermore, from the non-detection of polarisation in the SE hotspot at 144~MHz, 
with a degree of polarisation $<0.35$\%, we can place a lower limit on the Faraday 
depolarisation at this location of $\sigmaRM\sim0.25$\rad, based 
on comparison with the NVSS degree of polarisation of $\sim$5\% at this location. 
 
From inspection of the VLASS QL image at 3~GHz, the physical extent of the NW hotspot ($\sim$2.4\arcsec) 
is smaller compared to the SE lobe region (of order 20\arcsec~in size) and thus less affected by 
depolarisation caused by RM variations within the synthesised beam at 144~MHz. Since the amount of depolarisation scales roughly as the 
square-root of the number of Faraday rotation cells, this could reasonably explain the difference 
in the observed depolarisation between the lobes.  
However, the enhanced depolarisation at the location of the SE hotspot is more difficult to explain 
and may indicate a significant interaction between the hotspot/lobe magnetic field and the ambient 
medium. This warrants further investigation with more sensitive observations at low frequencies. 

Overall, given the small amount of observed Faraday depolarisation, it is important to consider the accuracy 
of the correction for Faraday rotation from the ionosphere. 
\cite{vaneck2018} estimate a residual error in the ionosphere RM correction between observations of 0.05\rad. As the ionosphere RM corrections across an observation (i.e.~8 hours) are linearly interpolated in time between direct estimates every 2 hours, a rough estimate can be made for the residual error within the observation of $\sim$$0.05\sqrt{4}\sim0.1$\rad. This means that most (or all) of the observed depolarisation in the NW hotspot is possibly due to residual errors in the ionospheric RM correction. 
However, the difference in depolarisation between the NW hotspot and SE lobe cannot be explained by ionosphere RM errors. 
Therefore, a $\sigmaRM$ of at least $\sim$0.1\rad~in the SE lobe can be considered astrophysically meaningful. 
This is comparable to the RM variations across the SE lobe of $\sim0.1$\rad~found in Section~\ref{sec:RM}. 

\begin{table}
\scriptsize{
\caption{Archival and measured flux densities, as well as the best-fit flux densities
(in the self-consistent, s.c., fits) for the north-west and south-east lobes of \grg.}
\begin{tabular*}{91mm}{lrclrcl}
\hline
\hline
       &$<$----------& N-lobe   &-------$>$&$<$----------& S-lobe& -------$>$ \\
Freq.  &  Entire Lobe & Hotspots & s.c. fit &    Entire Lobe & Hotspots & s.c. fit \\
(MHz)  &  [mJy]      &  [mJy]   &   [mJy]  &  [mJy]      &  [mJy]   &   [mJy]  \\
 (1)   &   (2)       &   (3)    &    (4)   &   (5)       &   (6)    &    (7)   \\
\hline
\\
143.6$^{(9)}$   & 403$\pm$40 &151$\pm$21 & 356.6 & 378$\pm$40 & 132$\pm$25 & 345.3 \\
151$^{(1)}$     &  350$\pm$52 &               & 344.4 &   320$\pm$52 &                & 333.3 \\
151$^{(2)}$     &  375$\pm$32 &               & 344.4 &   302$\pm$31 &                & 333.3 \\
325$^{(3)}$     &  177$\pm$36 &               & 193.0 &   149$\pm$36 &                & 185.1 \\
325$^{(9)}$     & 154$\pm$58 &               & 193.0 & 153$\pm$58 &                & 185.1 \\
408$^{(4)}$     &  160$\pm$40 &               & 160.6 &   145$\pm$34 &                & 153.2 \\
1400$^{(5)}$    &   59$\pm$4  &               &  55.9 &  50$\pm$2   &                &  51.0 \\
1400$^{(9)}$    & 55$\pm$19 & 36$\pm$4  &  55.9 &  47$\pm$19 &  33$\pm$5  &  51.0 \\
2980$^{(7)}$    &             & 21$\pm$3  &       &                              &  20$\pm$3 \\
4850$^{(6)}$    &   21$\pm$4  &               &  18.2 &  18.4$\pm$4  &                &  15.6 \\
15200$^{(8)}$   & (5.2$\pm$2)  &  5.2$\pm$1    &   6.3 & (6.6$\pm$2)  &   6.6$\pm$1    &   5.1 \\
\\
\hline
\label{tab:fluxes}
\end{tabular*}\\
{\bf References.} (1) 6C3 \citep{hales1990}; (2) 7Cn \citep{riley1999}; (3) WENSS \citep{rengelink1997};
(4) B3.3 \citep{pedani1999}; (5) NVSS \citep{condon1998}; (6) GB6 \citep{gregory1996}; (7) VLASS (Lacy et al.~in prep.);
(8) 9Cc \citep{waldram2010}; (9) this paper.
}
\end{table}

\begin{figure}
\begin{center}
\includegraphics[width=1\columnwidth,clip=true,trim=0.5cm 3.0cm 2.0cm 1.5cm]{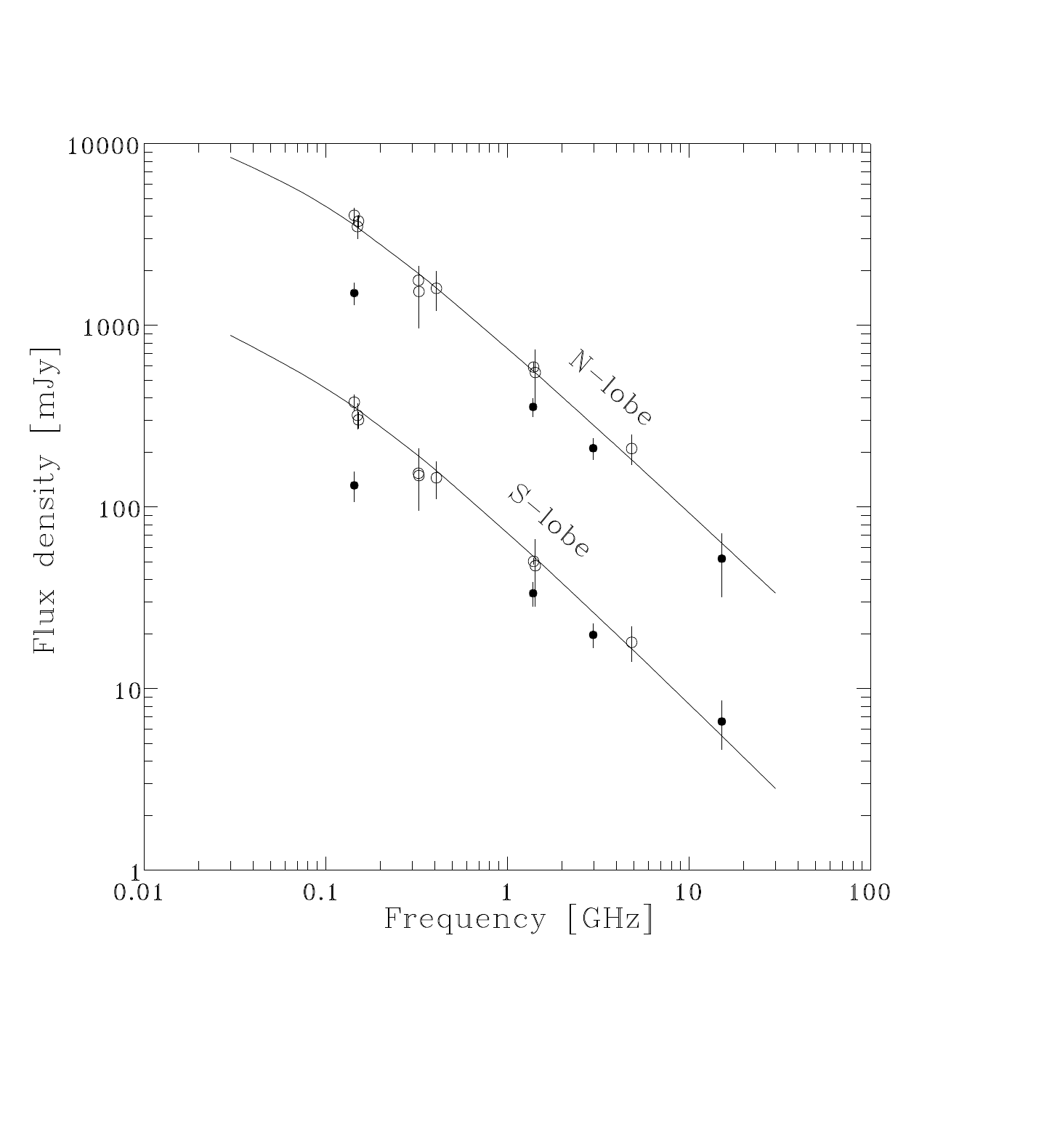}
\caption{DYNAGE fits (solid lines) to the total intensity spectra
of the north-west and south-east lobes (open circles), and the
spectral points of the hotspot regions (filled dots; not used in the fits).
Note that the north-west lobe flux density scale is shifted one
decade up in relation to the given ordinate scale. }
\label{fig:dynage}
\end{center}
\end{figure}

\subsection{Dynamical modelling}\label{sec:dyno}
In order to decouple the properties of the electron density and magnetic field along the line of sight 
in the measured Faraday rotation and depolarisation, additional information is required on the 
the physical characteristics of \grg~(i.e.~the magnetic field strength of the emission region) and the 
properties of its surrounding environment (i.e.~the ambient gas density). 
These properties can be estimated through dynamical modelling of the radio lobes, while simultaneously 
accounting for energy losses of relativistic particles (electrons and positrons) injected into the expanding lobes by the relativistic jets 
\citep[e.g.][and references therein]{machalski2011,machalski2016}. 
This is important because we lack X-ray data that could constrain the properties of the external medium \citep[e.g.][]{ineson2017}
and/or the magnetic field strength of the hotspot and lobes, without the need for the assumption of 
equipartition between the radiating particles and magnetic field \citep[e.g.][]{mingo2017}. 
Therefore, here we apply the evolutionary DYNAGE code of \cite{machalski2007} to the radio lobes of \grg, primarily 
to obtain an estimate of the external gas density, as well as estimates for the magnetic field strength of 
the lobes. 
The fitting procedure is performed separately for each lobe using the
observational data given in Section~\ref{sec:stokesi}, together with the radio
luminosities calculated from the flux densities listed in Table~\ref{tab:fluxes}. 
The input model parameters that are assumed are given in Table~\ref{tab:input}.

Characteristic of almost all FRII sources is a modest asymmetry in
the length and radio luminosity of the lobes. Therefore, as might be expected, the DYNAGE
results for the jet power $Q_{\rm j}$, the central density of the external medium $\rho_{0}$, and other physical
parameters can appear different for the two lobes of the same source. This aspect
has been analysed by \cite{machalski2009} and \cite{machalski2011} for a sample of thirty GRGs. 
While some of the differences were within the uncertainties of the fitted values 
for the model parameters, significant differences were possible in cases where the evolution of the magnetic field and/or 
various energy losses and acceleration processes of the relativistic particles are different at the hotspots of the opposite lobes.
Alternatively, such differences, especially in GRGs, may reflect different external
conditions well beyond the host galaxy and cluster/group environment.

Following \cite{machalski2009}, we averaged the values of $Q_{\rm j}$ and $\rho_{0}$ initially found in
the `independent solution' and treated them as fixed parameters in the 
`self-consistent' model, $\langle Q_{\rm j}\rangle$ and $\langle\rho_{0}\rangle$, respectively.
New values of the slope of the ambient density distribution ($\beta$) and the age ($t$) for the NW and SE lobes,
are denoted as $\beta_{\rm s.c.}$ and $t_{\rm s.c.}$ (Table~\ref{tab:output}). 
The DYNAGE fits to the observed data points are shown with solid lines in Figure~\ref{fig:dynage}.
Table~\ref{tab:output} presents the derived physical properties of the lobes, including a minimum-energy 
magnetic field strength in the lobes of $B_{\rm me}\sim1$~$\mu$G and an external 
density of $\sim2\times10^{-31}$~g~\cc~(i.e.~$n_{\rm e}\sim10^{-7}$\cc). This density is similar to 
the mean density of the Universe assuming half the baryons are in the WHIM \citep{machalski2011}, 
and implies that the radio lobes are likely propagating into a low-density region of the Universe. 

We also used the synchrotron minimum energy (equipartition) magnetic field formulation 
in \cite{worrallbirkinshaw2006} to estimate the lobe magnetic field strength.
From this we find an equipartition magnetic field strength that is 2.6 times higher than the 1~$\mu$G 
derived from the dynamical modelling (for $\gamma_{\rm min}=10$). 
When calculated in this manner the lobe equipartition field strength is usually found to be 
overestimated, by a factor of 2 to 3, compared to that found from X-ray Inverse Compton 
observations of lobes \citep[e.g.][]{ineson2017,mingo2017}. 
This highlights some of the uncertainties in the calculation of equipartition magnetic field strengths 
in radio galaxies \citep[e.g.][]{beckkrause2005, konar2008}.
Here we adopt the lobe magnetic field strength obtained from the dynamical modelling 
as it takes into account more physical effects, such as the jet power, adiabatic expansion and 
age of the lobes. 

\begin{table}
\scriptsize{
\caption{Dynamical modelling input model parameters}
\begin{tabular*}{75mm}{lcc}
\hline
\hline
Parameter    & Symbol   & Value  \\
\hspace{5mm}(1)&\hspace{-2mm} (2) & (3)   \\
\hline
\\

{\bf Set:} \\
Adiabatic index of the lobes' material          & $\Gamma_{\rm lb}$   & 4/3  \\
Adiabatic index of the ambient medium           & $\Gamma_{\rm x}$    & 5/3  \\
Adiabatic index of the lobes' magnetic field    & $\Gamma_{\rm B}$    & 4/3  \\
Minimum electron Lorentz factor (injected)      & $\gamma_{\rm min}$  & 1    \\
Maximum electron Lorentz factor (injected)      & $\gamma_{\rm max}$  & 10$^{7}$\\
Core radius of power-law \\
\hspace{15mm}ambient density distribution               & $a_{0}$             & 10\,kpc \\
Initial slope of power-law \\
\hspace{15mm}ambient density distribution               & $\beta$             & 1.5  \\
Thermal particles within the lobes              & $k$                 & 0    \\
Jet viewing angle                               & $\theta$            & 90$\degr$\\
\\
{\bf Free:} \\
Jet power                                       & $Q_{\rm j}$[erg s$^{-1}$] \\
External density at core radius                 & $\rho_{0}$[g cm$^{-3}$]   \\
Exponent of initial power-law energy  \\
\hspace{10mm}distribution of relativistic particles & $p=1+2\alpha_{\rm inj}$ \\
Source (lobe) age                               & $t$[Myr]\\ 
\hline
\label{tab:input}
\end{tabular*}
}
\end{table}

\begin{table}
\scriptsize{
\caption{Fitted values of the model free-parameters in the `self-consistent' dynamical modelling solution}
\begin{tabular*}{90mm}{lccc}
\hline
\hline
Parameter        & Symbol    &    Value    &    Value  \\
                 &           & for N-lobe  & for S-lobe \\
\hspace{5mm}(1)  &   (2)     &     (3)     &     (4)    \\
        \hline
        \\
Initial effective spectral index  & $\alpha_{\rm inj}$ & $-0.45$$\pm$0.05 & $-0.52$$\pm$0.03 \\
Source (lobe) age [Myr]           & $t_{\rm s.c}$      &   95$\pm$23   &   80$\pm$16    \\
Jet power [$\times 10^{45}$erg\,s$^{-1}$]    &$\langle Q_{\rm j}\rangle$ & 1.1$\pm$0.1 & 1.1$\pm$0.1\\
Core density [$\times 10^{-28}$g\,cm$^{-3}$] &$\langle\rho_{0}\rangle$ & 4.7$\pm$0.4 & 4.7$\pm$0.4 \\
Slope of ambient density 
                    distribution  & $\beta_{\rm s.c.}$      &   1.431    & 1.613 \\
External density [$\times 10^{-31}$g\,cm$^{-3}$] & $\rho(D)$  & 2.8$\pm$1.1 & 1.4$\pm$0.7 \\
Lobe pressure [$\times 10^{-14}$dyn\, cm$^{-2}$] & $p_{\rm lb}$& 3.0$\pm$0.1 & 3.1$\pm$0.1 \\
Minimum energy magnetic field [$\mu$G]    &$B_{\rm me}$        & 1.0$\pm$0.2  & 1.0$\pm$0.2  \\
Longitudinal expansion speed              &$v_{\rm h}/c$      &0.05$\pm$0.02 & 0.06$\pm$0.02 \\
\\
\hline
\label{tab:output}
\end{tabular*}
}
\end{table}

\section{Interpretation}
\label{sec:discuss}

The difference in the mean RM between the NW and SE lobes is $2.5\pm0.1$\rad. 
This may be due to variations in the Galactic RM (GRM) on scales of $\sim$11\arcmin, 
differences 
in the magnetoionic material of the intergalactic medium on large scales, 
and/or line-of-sight path length differences towards either lobe. 
The observed Faraday depolarisation of $\sigmaRM\sim0.1$\rad~associated with the SE 
lobe could be due to small scale fluctuations of the magnetic field in the local external 
medium and/or from Faraday rotation internal to the source. 
Constraining the likelihood of these possibilities requires some considerations of the expected variations 
in the GRM, knowledge of the geometry and physical properties of the radio lobes, and 
details of the environment surrounding the radio galaxy and in the foreground. 

\subsection{Galactic RM variations}
\label{sec:GRM}
The reconstruction of the GRM by \cite{oppermann2012,oppermann2015} gives 
$+14.8\pm4.5$\rad~across both the NW and SE lobe (the Galactic coordinates 
of \grg~are $l=128.46\degr$, $b=63.65\degr$). This is higher than the mean 
RMs of $+7.4$ and $+9.9$\rad~found for the NW and SE lobes, respectively. 
However, it should be kept in mind that the LoTSS RM values have been corrected for 
the time-variable ionosphere RM ($+1.6$ to $+1.9$\rad), while the catalogue from which the GRM map is mainly 
made \citep{taylor2009} does not have this correction applied. Thus, the RM 
of the NW and SE lobe are within the 1-sigma and 2-sigma errors in the GRM, respectively. 

The variation in the GRM map for three adjacent pixels (in the direction of the largest gradient) across the source is 
$\sim2.2$\rad~(on a scale of $\sim$1~deg).  
As the GRM map has a resolution of $\sim$1 degree, which is the typical spacing of extragalactic sources in the \cite{taylor2009} catalogue, it cannot be used to probe RM variations on smaller scales. 
The true GRM variation on smaller scales at this location is unknown, but RM structure function analyses for GRM variations at high Galactic latitudes have probed scales smaller than 1 degree in both observations  \citep[e.g.][]{mao2010, stil2011} and simulations \citep[e.g.][]{sunreich2009}. In particular, using the results from \cite{stil2011}, we find that GRM variations ranging from approximately 3\rad~to 13\rad~are possible on angular scales of $\sim$11\arcmin, depending on the highly uncertain slope of the RM structure function on angular scales less than 1 degree. 

Better estimates of the GRM are required to 
reliably remove the GRM and its variation across the extent of \grg. 

\subsection{Local environment RM contribution}
\label{sec:localRM}
The hot gas in rich groups and clusters is known to be magnetised from observations of synchrotron 
radio halos and relics, as well as Faraday rotation observations of embedded and background radio 
sources \citep[see][and references therein]{carillitaylor2002}. 
For radio galaxy lobes that have not expanded significantly beyond 
their host galaxy or cluster/group environment, the Laing-Garrington effect is often present 
\citep{laing1988,garrington1988,garringtonconway1991}. 
This is where the polarised emission from the counter-lobe travels through a greater amount of 
magnetoionic material and thus incurs a larger amount of Faraday depolarisation. 
However, as the lobes of \grg~are expected to be orientated close to the plane of the sky and 
extend well outside the influence of the group/cluster environment, 
the Laing-Garrington effect is not expected to be strong \citep[e.g.][]{laingbridle2014}.
Additionally, if the faint collimated emission SE of the host is indeed a jet, then the larger 
amount of depolarisation towards the SE lobe is opposite to that expected for 
the Laing-Garrington effect. 

Models of the variations in RM across radio galaxies in groups and clusters are typically constructed assuming turbulent magnetic field fluctuations over 
a range of scales embedded in a spherically-symmetric gas halo whose radial density profile is derived from X-ray 
observations \citep[e.g.][]{guidetti2008}. For \grg~we do not have X-ray data to constrain the properties of the hot gas environment, although 
it is likely that the red-sequence host galaxy is close to the centre of a poor cluster \citep{hao2010}. 
Therefore, we attempt to estimate the required density and field strength to self-consistently explain the mean RM and depolarisation \citep[e.g.][]{murgia2004}, for a single-scale model of a randomly orientated field structure \citep{felten1996}. 
In reality, the magnetic field will fluctuate on a range of scales, from an inner scale to an outer scale \citep{ensslinvogt2003}, but a single-scale model can provide a reasonable approximation to the RM variations if the scale length is interpreted as the correlation length of the magnetic field \citep[see][section 4.4 for details]{murgia2004}. 

An appropriate gas density profile, $n(r)$, for a galaxy group or cluster is a ``beta-profile'', where $n(r)=n_0(1+r^2/r_c^2)^{-3\beta/2}$. We assume that the magnetic field strength scales linearly with the gas density, $B(r)=B_0 n(r)/n_0$, where $B_0$ is the central magnetic field strength \citep[e.g.][]{dolag2001,laing2006,vacca2012, govoni2017}. Values of $n_0\sim10^{-3}$\cc, $r_c\sim100$~kpc and $\beta\sim0.5$ are not unreasonable for a poor cluster \citep[e.g.][]{laing2008, bonafede2010, guidetti2012}. 
The choice of these parameters is arbitrary given our limited information about the environment of the host galaxy (Section~\ref{sec:introduction}) but we use them simply as a plausible example. Following \citet[][eqn.~15]{murgia2004}, we find a Faraday dispersion 
of $\sigmaRM\sim0.1$\rad~at $r\sim1.5$~Mpc requires $B_0\sim5$~$\mu$G with a magnetic field correlation length of $\sim$25~kpc.
This implies an ambient density of $\sim$$1.7\times10^{-5}$\cc~and field strength $B\sim0.09$~$\mu$G at the location of the hotspots.\footnote{ For comparison, using a simple model 
with a constant electron number density of $n_e\sim10^{-5}$\cc~and constant magnetic field strength of $B_{||}\sim0.1$~$\mu$G, 
with a magnetic field reversal scale of $l\sim20$~kpc over a total path length of $L\sim1$~Mpc 
gives $\sigmaRM\sim0.81n_e\,B_{||}\,\sqrt{l\,L} \sim 0.1$\rad. } 
Using these values and a large outer scale for the magnetic field fluctuations of 500~kpc \citep{vacca2010} gives a mean $|$RM$|$ of $\sim$0.4\rad. Therefore, while we can reasonably explain $\sigmaRM\sim0.1$\rad~at $r\sim1.5$~Mpc, we cannot self-consistently explain the large mean RM excess of $\sim$2.5\rad, even for a large outer scale of turbulence in the magnetic field power spectrum \citep[][]{ensslinvogt2003,murgia2004}.
Note that the outer scale is mainly responsible for the observed mean RM and the inner scale for the value of $\sigmaRM$. 
We used a large outer scale here to show that this model cannot self-consistently explain both $\sigmaRM$ and the mean RM. 

Draping of the ambient field in addition to compression of the ambient magnetoionic gas could enhance the mean RM near the surface of the lobes \citep{guidetti2011, guidetti2012}, and may also help explain the higher depolarisation of $\sigmaRM\gtrsim0.15$\rad~at the location of the SE hotspot. 
Enhancements in the field strength and gas density by factors of 4 over a path length of $\sim$50~kpc outside the lobes could produce an additional $|$RM$|$ of $\sim$0.5\rad. More sensitive observations at high angular resolution are required to determine if such ordered field structures are indeed present. 

We note that the external gas density used here is two orders of magnitude higher than estimated from the dynamical modelling. This means that either the observed depolarisation does not occur in the external medium local to the source or that the dynamical modelling is severely underestimating the external density. Such low density gas may be challenging to detect in X-rays, but extrapolation of an X-ray profile from the inner region would be very instructive. In general, comparison with simulations of the propagation of large scale jets within a realistic cosmological environment may provide the best avenue for progress in this area \citep[e.g.][]{huarte2011, hardcastlekrause2014,  turnershabala2015, english2016, vazza2017}. 

\subsection{Internal Faraday depolarisation}
Our observations are insensitive to polarised emission from RM structures broader than 
$\sim1$\rad~(Section~\ref{sec:rmspecs}). Therefore, the large amounts 
of internal Faraday rotation required to explain the mean RM excess are ruled out. 
However, it is worth considering if the small amount of Faraday depolarisation ($\sigmaRM\sim0.1$\rad) 
can be explained by Faraday rotating material mixed with the synchrotron emitting material in the lobes. 

One of the most commonly used magnetic field models for the lobes of extragalactic sources is one where 
the field is highly tangled on small scales, with the observed appreciable degrees of polarisation produced due to 
stretching and compression \citep{laing1980}. 
Given the equipartition magnetic field strength of $\sim1$~$\mu$G within the lobes (Section~\ref{sec:dyno}), 
and as an illustrative example, we choose a thermal gas density internal to the lobes of $n_{\rm e}\sim 10^{-5}$\cc, 
with 500 field reversals through a lobe depth of $\sim$500~kpc, to produce 
$\sigmaRM\sim0.1$\rad~(using Eqn.~\ref{eqn:rm} and assuming $B_{||}=B/\sqrt{3}$). 
Observations at even lower frequencies would be required to resolve a Faraday depth width of 0.1\rad~in 
the Faraday spectrum (e.g.~using LOFAR observations down to at least 30~MHz, in combination with 
the data in this paper). In addition, broadband polarisation modelling would be needed to 
distinguish between internal and external Faraday depolarisation scenarios \citep[e.g.][]{anderson2018, osullivanlenc2018}. 
Using the LOFAR international baselines to obtain sub-arcsecond resolution would further enhance 
the ability to isolate different contributions by resolving the external RM variations across the emission region.

For now, we can assess the likelihood of this scenario in terms of the implied energetics. 
For expected internal thermal gas temperatures of $\gtrsim$10~keV \citep{gitti2007}, the lobe thermal gas pressure is 
$p_{\rm th}\sim 2n_{\rm e}kT\sim 3\times10^{-13}$~dyn~cm$^{-2}$, which is an order of magnitude 
larger than the pressure from the synchrotron-emitting plasma in the lobes ($p_{\rm lb}$ in Table~\ref{tab:output}). 
This is inconsistent with expectations from studies of other FRII lobes \citep{croston2005,ineson2017}, 
and thus unlikely, unless the internal thermal gas is much cooler than assumed here. 

\subsection{RM contribution from large-scale structure}

Significant asymmetries in the magnetoionic material in the foreground IGM, far from the local source environment, 
could also contribute to the observed mean RM difference between the lobes. Such variations 
could be caused by the magnetised component of the large scale structure (LSS) at low redshift, as  
\cite{ryu2008}, \cite{choryu2009} and \cite{akahoriryu2010} predict a root-mean-square RM ($\rm RM_{\rm rms}$) through LSS 
filaments of order 1\rad. In our case, the polarised emission of one lobe needs to pass through 
more foreground filaments than the other to explain the observed RM difference of 2.5\rad. 
Therefore, information is required on the location of LSS filaments with respect to the lines of sight 
probed by the polarised emission from the lobes of \grg. 

\subsubsection{Location of large scale structure filaments}
\label{sec:filaments}
\begin{figure}
\begin{center}
\includegraphics[width=1\columnwidth,clip=true,trim=0.0cm 0.0cm 0.4cm 0.0cm]{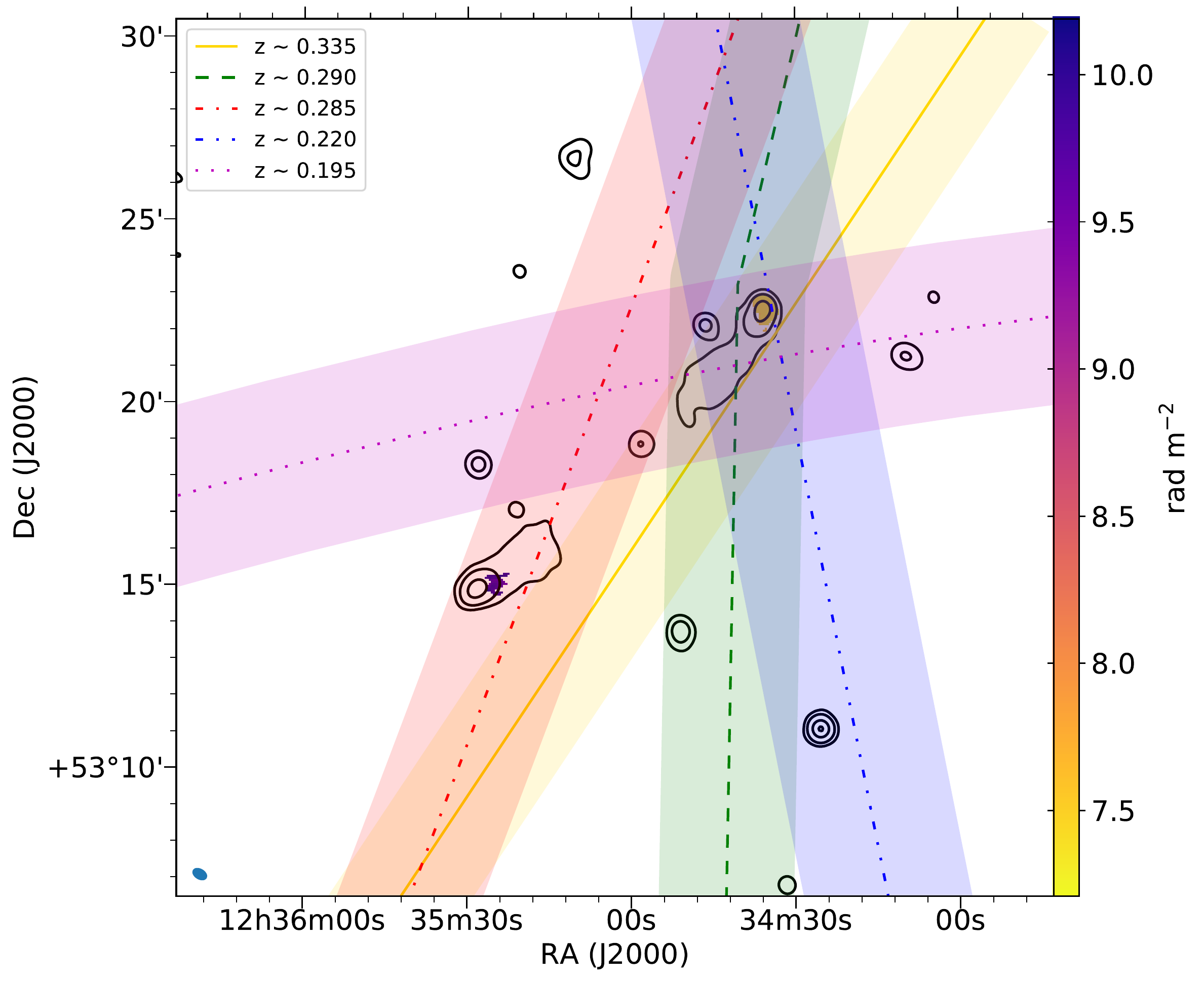}
\caption{Location of foreground large-scale-structure filaments (lines) 
in relation to the background radio galaxy (contours) and its Faraday rotation measure (colour scale), as described in Fig.~\ref{fig:RMfpol}. 
The width of the lines corresponds to $\sim$1~Mpc at the redshift of the 
filament. }
\label{fig:filaments}
\end{center}
\end{figure}

The catalogue of \cite{chen2015,chen2016} provides a cosmic filament 
reconstruction from the SDSS data for 130 redshift slices in the range $0.05 < z < 0.7$. In Figure~\ref{fig:filaments}, 
we plot the location of the filaments that are in the foreground of \grg~(i.e.~at $z<0.34$). There are five 
filaments identified in different foreground redshift slices that pass through the field. We assign a thickness 
of 1~Mpc to each filament \citep{vazza2015} to determine which filaments most likely intersect lines of sight towards the 
polarized lobes (Figure~\ref{fig:filaments}). For a thickness of 1~Mpc, there are 
four filaments that cover the NW lobe and one filament that covers the SE lobe. Therefore, we estimate 
that there is an excess of three filaments covering the NW lobe. Considering different filament thicknesses 
results in different numbers of filaments covering each lobe, with an excess of filaments covering the 
NW lobe remaining for filaments up to a thickness of $\sim$3.8~Mpc 
(i.e.~the thickness above which the same number of filaments cover both lobes). 
In light of this result, we consider if 
the RM difference between the lobes can be explained by magnetised gas in these filaments. 
We note that there is no evidence of an individual intervening galaxy in the SDSS images that could 
explain the RM difference. 

\subsubsection{Magnetic field stength in filaments}\label{sec:meanbfield}
To explain the RM difference between the lobes, an RM excess of $-2.5$~\rad~must be provided 
by the three extra filaments covering the NW lobe.  
Simulations suggest that the electron number density of LSS filaments can vary from $10^{-6}$ to $10^{-4}$\cc~
\citep{cenostriker2006,ryu2008,choryu2009,akahoriryu2010,vazza2015}, thus we adopt a mean 
electron density of $10^{-5}$\cc. 
\cite{akahoriryu2011} found a peak in the RM power spectrum, due to their simulated IGMF in filaments, 
on scales corresponding to a proper length of $\sim$3~Mpc, which they expect to correspond to the 
typical line-of-sight path through LSS filaments. 
Therefore, using a path length ($L$) of 3~Mpc and a coherence length ($l$) of 300~kpc \citep{choryu2009} 
leads to a magnetic field strength in the filaments ($B_{\rm LSS}$) of approximately
\begin{equation}\label{eqn:Brms}
B_{\rm LSS} \sim 0.3 \left( \frac{n_{\rm e}}{10^{-5}\,{\rm cm^{-3}}} \right)^{-1} \left( \frac{L}{3(3\,{\rm Mpc})} \frac{l}{300\,{\rm kpc}} \right)^{-1/2} \mu{\rm G},
\end{equation}
for $B_{\rm ||}=B_{\rm LSS}/\sqrt{3}$. 
This estimate of the density-weighted IGMF strength of $\sim0.3$~$\mu$G has significant uncertainty given our limited knowledge of the particle number density of the gas in these filaments, as well as the observationally unconstrained coherence length of the field and the path length though each filament.  
Furthermore, this estimate cannot be treated as an upper limit as a large Galactic RM variation across the source (Section~\ref{sec:GRM}) could make the difference in RM between the lobes even larger (since the RM can be positive or negative). 
Furthermore, much larger RM variations are observed across radio relics which cannot be explained by Galactic RM variations, indicating the presence of large scale ordered fields in the outskirts of galaxy clusters \citep[e.g.][]{kierdorf2017,loi2017}. 

Therefore, a better approach may be to compare directly with cosmological simulations of the RM 
contribution from such LSS filaments. 
These simulations suggest that the magnetic field strength in filaments could range 
somewhere from $\sim$1 to 100~nG \citep[e.g.][]{vazza2015}.
Early hydrodynamic simulations by \cite{ryu2008} used a prescription to produce magnetic fields from 
the kinetic energy of turbulent gas flows (guided by expectations from small-scale magnetic dynamo simulations), 
which produced average IGMF strengths of $\sim10$~nG.
Subsequent work by \cite{choryu2009} and \cite{akahoriryu2010,akahoriryu2011}, using the results of these simulations, 
provided estimates of the ``typical'' RM contribution from LSS filaments. 
The most relevant number for Faraday 
rotation is the gas density ($\rho$) weighted average of the strength of the magnetic field through the filaments, 
i.e.~$\langle (\rho B)^2 \rangle^{1/2} / \langle \rho^2 \rangle^{1/2} $, which gave a few $\times$~$0.1$~$\mu$G in 
the above simulations. From this, it was found that the root-mean-square RM (RM$_{\rm rms}$) 
through the filaments scales with the number of filaments ($N_{\rm f}$) as 
RM$_{\rm rms}\sim1.5N_{\rm f}^{1/2}$\rad, up to a saturation point that corresponds to 
$\sim$25 filaments for $z>1$. In the case of three filaments, the predicted RM$_{\rm rms}\sim2.6$\rad, which is consistent with our 
observations (where we have an RM difference of 2.5\rad~between only two lines of sight, in which one passes 
though three additional filaments). 
Therefore, it can be argued that our results are consistent with the 
expected Faraday rotation signature from an average magnetic field strength in LSS filaments of $\sim10$~nG. 

We further investigated the above findings by direct comparison with recent MHD cosmological simulations, 
as described in \cite{va14mhd}.  
In particular, we analysed the 
RM distribution in the warm-hot gas simulated in a cosmic volume of $50^3\rm ~Mpc^3$, at a spatial resolution of $20~\rm kpc$ (comoving). To better compare with our observations, we generated a long integration cone for this volume, stacking several  randomly oriented, mirrored replicas of the volume, covering the comoving distance out to $z=0.34$.  
In this way, we could measure the probability of having a contribution as large as 2.5\rad~from LSS filaments 
for the \grg~observations at $z=0.34$.  
We found that this occured in only 5\% of cases, for typical magnetisation values of $\sim$10 to 50~nG, amplified from an initial magnetic field strength of 1 nG, which was seeded at an early cosmological epoch and is in line with the upper limits given by the Planck satellite \citep[][]{PLANCK2015}. The probability was negligible for a significantly smaller seed field of 0.1~nG.  

Lower limits on the primordial field strength of $\sim$$10^{-16}$~G \citep{neronov2010} and $\sim$$10^{-20}$~G \citep{takahashi2013} imply that the true value may indeed be much lower. 
However, this is not the only possible scenario, as the LSS can be magnetised by a more ``astrophysical'' mechanism, such as galaxy feedback \citep[e.g.][for a recent review]{va17cqg}, or produced by a more efficient dynamo amplification of primordial fields \citep[][]{ryu2008} than is found in current MHD simulations. 
Therefore, from comparison with the MHD simulations, we consider it unlikely that the true RM contribution 
from the IGMF is as large as 2.5\rad, and that the observed RM excess is possibly dominated by other contributions 
along the line of sight, such as small scale GRM variations (Section~\ref{sec:GRM}). 

\section{Conclusions}
\label{sec:conclusion}
We have presented a linear polarisation and Faraday rotation study of a giant FRII 
radio galaxy, \grg, using data from the LOFAR Two-Metre Sky Survey \citep{shimwell2018}. 
After obtaining the spectroscopic redshift of the host galaxy (SDSS~J123501.52$+$531755.0, $z=0.3448\pm0.003$), 
we find that the radio galaxy has a projected linear extent of 3.4~Mpc. 
Both lobes are detected in polarisation with a mean RM difference between the lobes of $2.5\pm0.1$\rad. 
Small amounts of Faraday depolarisation ($\sim0.1$\rad) are also detected. 
In the absence of direct tracers of the gas density on large scales, we employ dynamical 
modelling of the advancing hotspots to infer a particle number density of the ambient gas 
of $n_{\rm e}\sim10^{-7}$\cc. This implies that the radio galaxy is expanding into an underdense 
region of the Universe. 
However, explaining the observed Faraday depolarisation (that most likely occurs in the 
environment local to the source) requires $n_{\rm e}\sim10^{-5}$\cc~in combination with a turbulent 
magnetic field strength of $\sim$0.09~$\mu$G at a distance of $\sim$1.5~Mpc from the host galaxy. 
Therefore, either the dynamical modelling is underestimating the density of the external medium 
or the depolarisation does not occur in the local source environment. 
Simulations of the propagation of FRII jets to large scales within a realistic cosmological 
environment may help distinguish between these scenarios. 
In general, the estimated magnetic field strength is unable to account for the observed 
mean Faraday rotation difference of 2.5\rad~between the two lobes. 

Using a catalogue of large scale structure (LSS) filaments in the local universe derived from optical spectroscopic 
observations, we find an excess of 
filaments intersecting lines of sight towards the polarised emission of the NW lobe. 
Assuming that magnetised gas 
in these LSS filaments is responsible for the RM difference between the lobes, gives a density-weighted magnetic field 
strength of 0.3~$\mu$G (assuming $n_{\rm e}\sim10^{-5}$\cc, a line-of-sight path length through each filament of 3~Mpc, 
and a magnetic field coherence length of 300~kpc). 
However, we find that predictions from cosmological simulations of the RM contribution from LSS filaments 
gives a low probability ($\sim$5\%) for an RM contribution as large as 2.5\rad. This probability applies to the case 
of magnetic fields strengths in the LSS filaments of 10 to 50 nG, which are amplified from primordial magnetic fields 
close to current upper limits from the CMB of $\sim$1~nG (the probability decreases to $\sim$0\% for weaker fields).
Extrapolation of the observed variations in the Milky Way RM to 11\arcmin~scales (i.e.~the angular size of \grg) 
indicates that this likely contributes significantly to the mean RM difference, however, further observations are required 
to obtain better constraints. 

In the near future, large samples of RMs from radio galaxies with known redshifts will allow 
more advanced statistical analysis techniques to be used, such as RM structure function analyses \citep[e.g.][]{akahori2014} 
and cross-correlation with other tracers of LSS \citep[e.g.][]{stasyszyn2010, vernstrom2017, brown2017}. 
This will enable a better separation of 
the Faraday rotation due to our Galaxy \citep[e.g.][]{haverkorn2004,sunreich2009,mao2010,stil2011} from that due to the cosmic web, and put stronger constraints 
on the strength and structure of the intergalactic magnetic field.

\begin{acknowledgements}
This paper is based (in part) on data obtained with the International LOFAR
Telescope (ILT) under project codes LC2\_038 and LC3\_008. LOFAR \citep{vanhaarlem2013} is the Low
Frequency Array designed and constructed by ASTRON. It has observing, data
processing, and data storage facilities in several countries, that are owned by
various parties (each with their own funding sources), and that are collectively
operated by the ILT foundation under a joint scientific policy. The ILT resources
have benefitted from the following recent major funding sources: CNRS-INSU,
Observatoire de Paris and Universit\'e d'Orl\'eans, France; BMBF, MIWF-NRW, MPG,
Germany; Science Foundation Ireland (SFI), Department of Business, Enterprise and
Innovation (DBEI), Ireland; NWO, The Netherlands; The Science and Technology
Facilities Council, UK; Ministry of Science and Higher Education, Poland.
SPO and MB acknowledge financial support from the Deutsche Forschungsgemeinschaft (DFG) under grant BR2026/23. 
Part of this work was carried out on the Dutch national e-infrastructure with the support of the SURF Cooperative through grant e-infra 160022 \& 160152. The LOFAR software and dedicated reduction packages on https://github.com/apmechev/GRID\_LRT were deployed on the e-infrastructure by the LOFAR e-infragroup, consisting of J. B. R. Oonk (ASTRON \& Leiden Observatory), A. P. Mechev (Leiden Observatory) and T. Shimwell (ASTRON) with support from N. Danezi (SURFsara) and C. Schrijvers (SURFsara).  
This research has made use of data analysed using the University of
Hertfordshire high-performance computing facility
(\url{http://uhhpc.herts.ac.uk/}) and the LOFAR-UK computing facility
located at the University of Hertfordshire and supported by STFC
[ST/P000096/1]. 
This research made use of Astropy, a community-developed core Python package for astronomy \citep{astropy2013} hosted at http://www.astropy.org/, of Matplotlib \citep{hunter2007}, of APLpy \citep{aplpy2012}, an open-source astronomical plotting package for Python hosted at http://aplpy.github.com/, and of TOPCAT, an interactive graphical viewer and editor for tabular data \citep{taylor2005}. 
FV acknowledges financial support from the ERC Starting Grant "MAGCOW", no.714196, and the usage of e  usage of computational resources on the Piz-Daint supercluster at CSCS-ETHZ (Lugano, Switzerland) under project s701 and s805. 
Based on observations made with the Nordic Optical Telescope, operated by the Nordic Optical Telescope Scientific Association 
at the Observatorio del Roque de los Muchachos, La Palma, Spain, of the Instituto de Astrofisica de Canarias.
KEH and JPUF acknowledge support by a Project Grant (162948-051) from The Icelandic Research Fund. 
The Cosmic Dawn Center is funded by the DNRF.
RJvW acknowledges support from the ERC Advanced Investigator programme NewClusters 321271 and the VIDI research programme with project number 639.042.729, which is financed by the Netherlands Organisation for Scientific Research (NWO). 
HA benefited from grant DAIP \#66/2018 of Universidad de Guanajuato. KT is partially supported by JSPS KAKENHI Grant Number 16H05999 and 17H01110, MEXT KAKENHI Grant Number 15H05896, and Bilateral Joint Research Projects of JSPS. LKM acknowledges support from Oxford Hintze Centre for Astrophysical Surveys which is funded through generous support from the Hintze Family Charitable Foundation. This publication arises from research partly funded by the John Fell Oxford University Press (OUP) Research Fund. 
SPO thanks A.~G.~de Bruyn for stimulating discussions on the topic of this paper, and the referee for their helpful comments. 
\end{acknowledgements}

%
   \bibliographystyle{aa} 
   \bibliography{grgpol} 
%

\end{document}